# Inverted Channel Belts and Floodplain Clays to the East of Tempe Terra, Mars: Implications for Persistent Fluvial Activity on Early Mars

Zhenghao Liu[1,2], Yang Liu[1,3], Lu Pan[4], Jiannan Zhao[5], Edwin S. Kite[6], Yuchun Wu[1,2], and Yongliao Zou[1]

[1]State Key Laboratory of Space Weather, National Space Science Center, Chinese Academy of Sciences, Beijing 100190, China
[2]University of Chinese Academy of Science, Beijing 100049
[3]Center for Excellence in Comparative Planetology, Chinese Academy of Sciences, Hefei 230026, China
[4]University of Copenhagen, GLOBE Institute, Center for Star and Planet Formation
[5]Planetary Science Institute, China University of Geosciences, Wuhan 430074, China
[6]University of Chicago, Chicago, IL, 6063, USA

**Key Points**

- Inverted sinuous ridges and flood plain clays were found to the east of Tempe Terra on Mars
- The inverted ridges may represent exhumation of the channel belts and overbank deposits
- The findings imply prolong fluvial activity on early Mars, potentially for tens of thousands of years

**Abstract**

The climate on early Mars is one of the major unsolved problems. It is unclear whether early Mars was warm and wet or cold and icy. Morphological features on Mars such as sinuous ridges could provide critical constraints to address this issue. Here we investigate several sinuous ridges to the east of Tempe Terra, Mars and find they may have recorded persistent fluvial activity on early Mars. Our analysis indicates that these ridges may represent exhumation of the channel belts and overbank deposits formed from meander rivers over significant geologic time. Layered smectite-bearing minerals, distributed along the ridge flanks, could be detrital or authigenic floodplain clays. Our interpretation of the stratigraphic relationships indicates that the layered smectite-bearing materials lie between channel belt deposits, supporting the floodplain interpretation. Our results suggest that a persistent warming event has persisted for a geologically significant interval (>1500 yr) during the Noachian period of Mars.

**Plain Language Summary**

It is not clear whether Mars in ancient time was warm and wet or cold and icy. Studying the morphology and the mineralogy of Martian terrains can us help understand the climate history of Mars. In this study, we found several sinuous ridges, a positive-relief landform, in the eastern area of Tempe Terra on Mars. Using orbital remote sensing data sets, we discovered clay-bearing hydrated minerals along the slope of the ridges and also analyzed the detailed morphology of these ridges. The sinuous ridges identified in this area could be inverted channel belts. Channel belt form when the river meandering over a relatively flat area floods the area and deposits sediments on it. The exhumation of channel belts and overbank deposits would form inverted ridges. The clay-bearing hydrated minerals could have been transported to the deposition sites or formed in place. These results suggest that a warming event that supports persistent fluvial activity may have lasted over at least thousands to tens of thousands of years on early Mars.

**Introduction**

The nature and evolution of the climate of early Mars is a fascinating and fundamental, yet unsolved problem. Widely distributed clay minerals, as well as geomorphological features such as valley networks, alluvial fans, lake basins, and deltas, suggest the presence of liquid water and a persistent warm and wet climate on early Mars [*Craddock and Howard*, 2002; *Howard*, 2007; *Howard et al.*, 2005; *Mischna et al.*, 2013; *Ramirez and Craddock*, 2018]. However, global climate modelling suggest Mars could have been cold and icy in the Noachian with only transient warm and wet periods [*Wordsworth et al.*, 2015; *Palumbo and Head*, 2020], which could also explain the distribution of sedimentary rocks and other geomorphic observations [*Kite et al.*, 2013a; *Wordsworth*, 2016]. Geomorphic and geochemical analysis of Martian terrains could provide critical information to test and constrain global climate models for early Mars. Of particular interest among Mars' geomorphological landforms are sinuous ridges. Sinuous ridges are widely distributed on the surface of Mars, showing a variety of ridge cross section types and a length of tens to hundreds of kilometers [*Burr et al.*, 2009; *Hiesinger and Head*, 2002; *Williams et al.*, 2013; *Williams et al.*, 2009]. Although some sinuous ridges on Mars were considered to be eskers [*Butcher* et al. 2016], many sinuous ridges may be composed of strata from fluvial depositional systems and could have recorded a prolonged aqueous history over millions of years [*Hayden et al.*, 2019]. Therefore, the analysis of sinuous ridges could provide important insight on climate conditions on early Mars.

Many sinuous ridges have been identified and analyzed on Mars [*Banks et al.*, 2009; *Williams et al.*, 2013], however, it is much less common for hydrated minerals to be reported in association with sinuous ridges, except one case in Miyamoto crater, near the Opportunity landing site [*Marzo et al.*, 2009; *Newsom et al.*, 2010; *Wiseman et al.*, 2008]. In many terrains that contain valley networks, topographically inverted ridges were interpreted to be channel-fill deposits that have been exhumed to form inverted river channels due to differential erosion. These inverted paleochannels have been found in Arabia Terra [*Davis et al.*, 2019], Aeolis Dorsa/Medusae Fossae Formation [*Lefort et al.*, 2012; *Williams et al.*, 2013] and Miyamoto crater [*Newsom et al.*, 2010], among other sites, and provide evidence for prolonged fluvial activity in these areas. Alternatively, these sinuous ridges are interpreted to be inverted deposits which may represent exhumation of the channel belts and overbank deposits formed from meandering rivers over significant geologic time [*DiBiase et al.* 2013; *Hayden et al.*, 2019]. If this is the case, the sequence of fluvial floodplain and channel-belt deposits contained in these ridges may preserve markers of chemical alteration to form hydrated minerals (either in-situ or in upstream soils). However, it remains unclear whether channel-belt deposits and related floodplain clays exist on Mars.

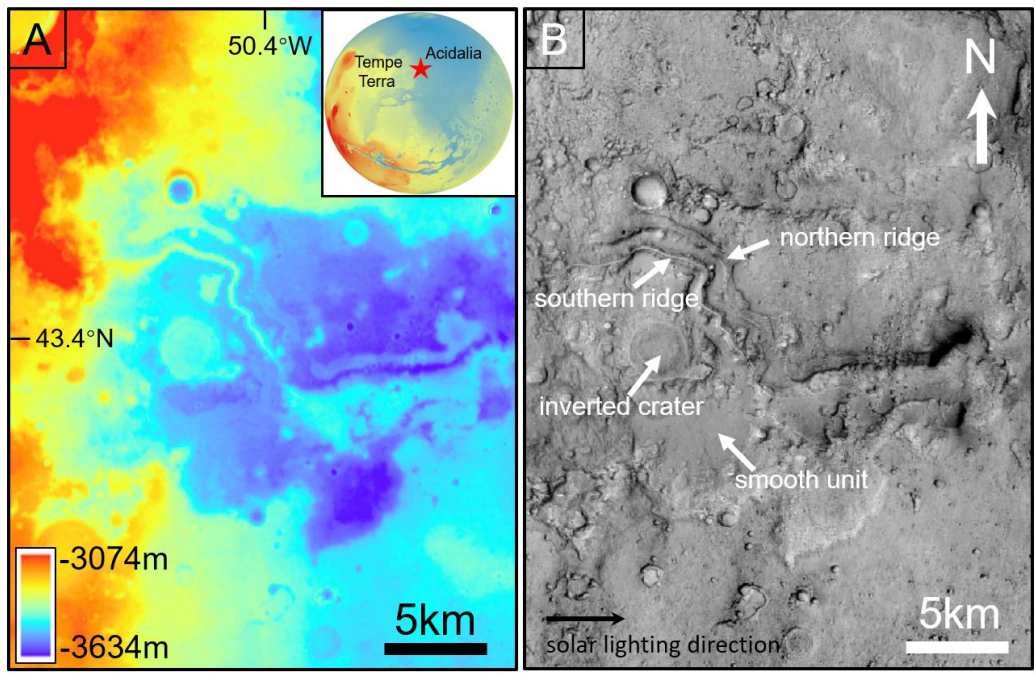

**Figure 1.** (a) CTX DEM (J19_052275_2237_XN_43N050W and J20_052552_2237_XN_43N050W) of the study area. The inset on the top right is the Mars Orbiter Laser Altimeter (MOLA) global map where the study region is indicated by red star. Paralleled inverted sinuous ridges are evident in a depression. (b) Geological context of the study area shown in the CTX image (J19_052275_2237_XN_43N050W) which is same as the image in (a). Lighting direction is the left.

Here we report a detailed analysis of sinuous ridges to the east of Tempe Terra, located at the northernmost highland on Mars. These sinuous ridges show distinct geomorphology and are consistent with inverted channel belts and overbank deposits. The clay minerals associated with these inverted ridges are analyzed and their formation mechanisms are discussed. Finally, the inferred climatic conditions to form these sinuous ridges and clay minerals on early Mars are presented.

**Data Sets and Geologic Settings**

We used data from Mars Reconnaissance Orbiter (MRO) instruments, including the Context Camera (CTX), the High-Resolution Imaging Science Experiment (HiRISE), and the Compact Reconnaissance Imaging Spectrometer for Mars (CRISM). The CTX images (spatial resolution of 5-6.5 m/pixel) were used to examine the geological context of ridges. HiRISE images (resolution up to 0.3 m/pixel) [*McEwen et al.*, 2007], were used to examine the geomorphic details. Digital elevation models (DEM) were constructed from both CTX and HiRISE stereopairs, and were also utilized for quantitative analysis on the topography. CRISM data with targeted mode (18-36 m/pixel, 362-3920 nm at 6.55 nm/channel) were processed with the volcano scan correction method and used to identify the type and distribution of minerals in the study area [*Murchie et al.*, 2007; *McGuire et al.*, 2009].

The sinuous ridges identified in this study are located to the east of Tempe Terra at the dichotomy boundary of Martian highland and lowland plains (**Figure 1**), which have been previously described by *Pan* and *Ehlmann* [2014]. Tempe Terra is the northernmost exposure of ancient heavily cratered highland terrain and has a complex history [*Tanaka et al.*, 2014]. Tempe Terra grades into Acidalia Planitia to the east, connecting the old Noachian terrain to the smooth northern plains. Our study region lies in a stratigraphy context at the transition from the highland material in Tempe Terra to the lowlands of Acidalia Planitia (**Figure 1**), similar to the knobby plains unit in Acidalia near Arabia Terra [*Pan* and *Ehlmann*, 2014].

**Results**

The sinuous ridges consisting of two parallel ridges are evident in the CTX DEM map (**Figure 1a**). The geologic context of these ridges is shown in **Figure 1b**. The ridges are 30-35 km long, extending down slope to the east at the boundary between the highland terrain and the lowland plains. The ridges are 50-60 m in height with an average flank slope between 0.15 and 0.25, superposing relatively old terrain (middle to late Noachian, [*Tanaka et al.*, 2014]). An inverted impact crater is observable to the southwest of the southern ridge, indicating the area has experienced extensive exhumation and denudation (**Figure 1b**). The smooth unit to the southeast of the inverted crater represents a resurfacing event that may have filled the previous topography with lava or sediments (**Figure 1b**).

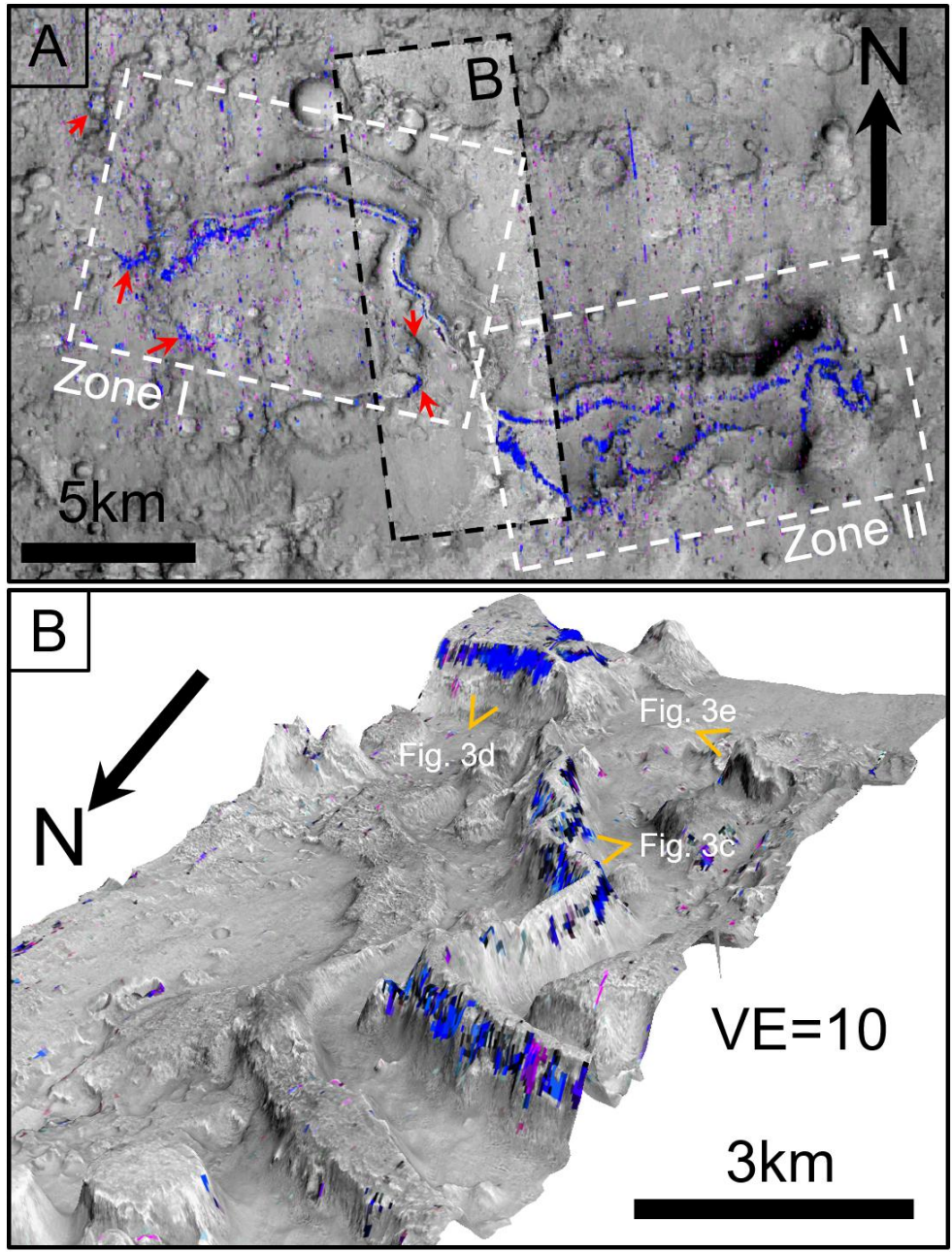

**Figure 2.** (a) CRISM parameter map overlain on the CTX (J20_052552_2237_XN_43N050W) and HiRISE (ESP_052275_2235_RED) images. Areas with hydrated mineral signatures are indicated by blue to purple color. The ridges were divided into Zone I and II based on the different widths of the southern ridge. The hydrated minerals are distributed on both sides of the ridge flanks. These materials are also exposed in various settings close to the ridge, indicated by the red arrows. (b) HiRISE DTM image generated from HiRISE stereopair ESP_052275_2235_RED/ESP_052552_2235_RED. The hydrated minerals are distributed on both sides of the ridge flanks. Orange symbols show the locations and the approximate viewing directions

for the perspective HiRISE views in Figures 3c-d. A vertical exaggeration (VE) of 10 was used in the 3D visualization of the DTM.

The mineralogy associated with the sinuous ridges was investigated using CRISM data. To map the distribution of hydrated minerals, we applied the spectral parameters of *Viviano-Beck et al.* [2014] using the band depths at 1.9 μm (BD1900R2, blue), 2.1 μm (BD2100_2, green), and the convexity at 2.29 μm due to the absorption at 1.9/2.1 μm & 2.4 μm (SINDEX2, red). Consistent with an earlier study [*Pan and Ehlmann*, 2014], we found the hydrated minerals are continuously distributed along both sides of the southern ridge as well as in various settings close to the ridge (**Figure 2a**). The CRISM parameter maps were co-registered with the HiRISE and CTX digital terrain model (DTM) using manually picked ground control points to map the stratigraphic relationships of hydrated minerals and other units (**Figure 2b**), which clearly shows that the hydrated mineral unit is located at the upper part of the slope, parallel to the capping unit of the ridge. The hydrated minerals show a combination of absorptions at 1.4 μm, 1.9 μm, and 2.3 μm, which are mostly consistent with Fe/Mg smectites (**Figure 3**). Fe/Mg smectites show typical absorptions at 1.4 μm due to $H_2O$ and metal-OH vibrations, at 1.9 μm due to $H_2O$, and at 2.3 μm due to Fe/Mg-OH. The Fe and Mg endmembers of the smectite group have slightly different band centers around 1.4 and 2.3 μm [*Ehlmann et al.*, 2009]. The CRISM spectra are mostly centered at 1.39 and 2.31 μm (**Figure 3**), indicating the materials are most likely a Mg-rich smectite such as saponite.

We show the HiRISE DTM perspective views of representative clay exposures on the flank of the sinuous ridges at Zone I (**Figure 3c**), Zone II (**Figure 3d**) and on the slope of a small butte to the south of ridges (**Figure 3e**). At Zone I, fine-grained layers with different thicknesses below the light-toned caprock (LTC) are visible, and the clay-bearing bedrock is mostly associated with the mid-toned and the lower dark-toned layers (MTL & LDTL) (**Figure 3c**). The clay mineral at the lower and thick mid-toned layer (LMTL) could correspond to clay-bearing clasts that could have rolled down from higher levels. The clay-bearing bedrock at Zone II and on the slope of the small butte to the south of the sinuous ridges is light-toned, and the texture of these outcrops is much rougher compared to Zone I (**Figure 3d** and **3e**). Eolian ripples and debris are evident immediately beneath these light-toned deposits, masking the contact with the underlying unit.

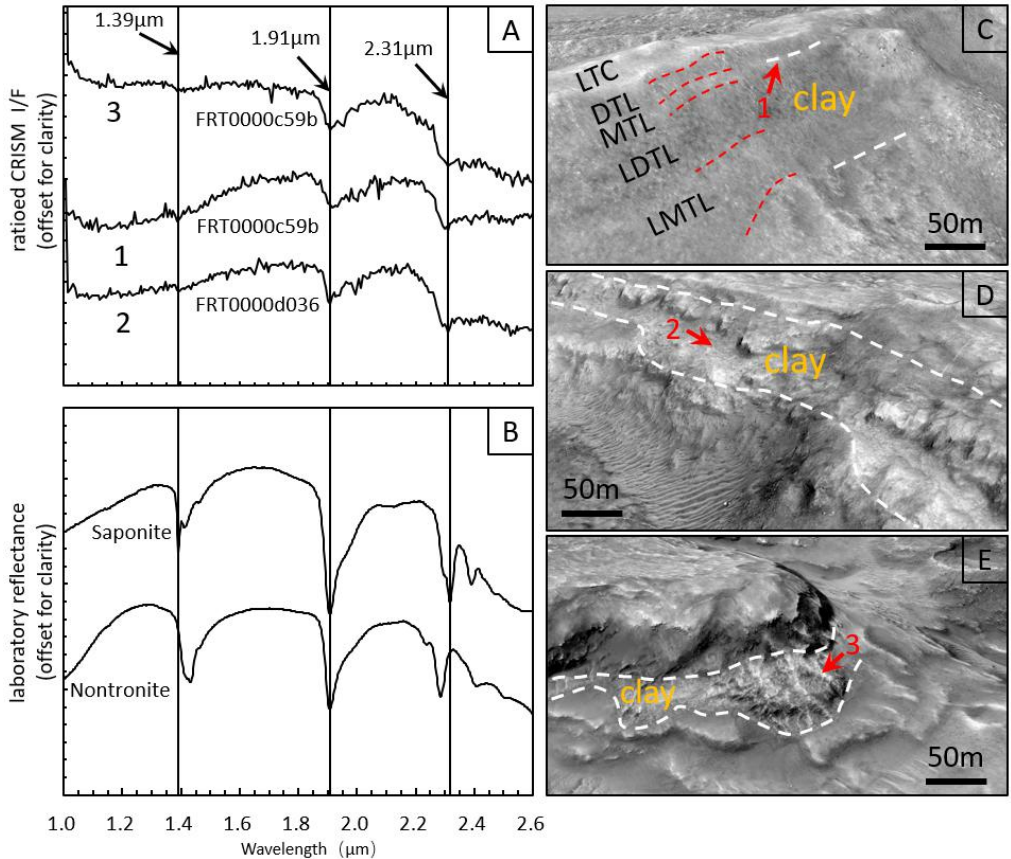

**Figure 3.** (a) Ratioed CRISM I/F spectra (spectra with hydrated mineral signatures ratioed over the bland spectra lacking spectral features in a similar column in the image) of Locations 1, 2, and 3. The locations of the spectra shown are indicated with red arrows in Figures c-e. The spectra show absorption bands at 1.91 and 2.31 μm, with weak absorption features at 1.4 μm. (b) Laboratory spectra of saponite and nontronite from Reflectance Experiment Laboratory (RELAB, a NASA multiuser spectroscopy facility) in Brown University. The band positions of hydrated minerals in our study area are mostly consistent with those of saponite. (c-e) HiRISE DTM perspective views of clay mineral-bearing exposures on the flank of the sinuous ridges at Zone I and Zone II and on the slope of a small butte to the south of the ridges, respectively. The locations of these DTMs are indicated in Figure 2b. LTD = light-toned caprock; DTL = dark-toned layer; MTL = mid-toned layer; LDTL = lower, dark-tone layer; LMTD = lower, mid-tone layer.

Interestingly, the smectites are found on the flanks of the southern ridge while no hydrated minerals were detected on the northern ridge. The topographic profiles along the two ridges and the channel show decreased elevation to the east (**Figure 4a**). The slope and width of the ridges along west to east direction are plotted in **Figure 4b**, indicating a steeper slope and smaller widths of the ridges at Zone I as compared to Zone II. The cross sections of the Zone I and Zone II of the two ridges are shown in **Figures 4c** and **4d**, respectively, both showing the southern ridge is much higher than the northern ridge. Because the horizontal offset between the ridges is only 1-2 km, the long-baseline slopes of the ridges are very gentle (average <0.5°, **Figure 4a**), and there is no evidence for faulting at this site, we believe topographic elevation is a proxy for stratigraphic position. In turn, this suggests that the caprock material of the northern ridge stratigraphically predates the materials of the southern ridge, and that the clay-bearing material in the southern ridge is stratigraphically encapsulated by episodes of deposition of material that formed sinuous-ridge caprock (**Figures 4c-d**). This pattern, with sinuous deposits of erosionally-resistant materials bracketing clay-bearing, fine-grained erosionally recessive material, is also seen in terrestrial river-and-floodplain deposits [*Bridge*, 2003]. This pattern has not previously been reported on Mars. It is possible that the northern ridge has similar clay mineral deposits beneath the capping unit, but that they are not visible due to differences in erosion and/or mass-wasting. Alternatively, clay may have never been formed at the location of the northern ridge. The topographic cross sections of Zone I and Zone II show that the ridges have flat-topped crests, and the surfaces of ridges for Zone II are rougher especially for the southern ridge due to cratering processes (**Figure 4c** and **4d**). The cross section also shows that the 20-m thick clay-bearing unit underlies a 4 m thick caprock unit for the southern ridge, whereas the northern ridge lacks such a clay-bearing unit. The exposed clay-bearing strata sit along the cliff-forming caprocks and ridges (**Figure 2b**), suggesting that significant lateral erosion has occurred [*DiBiase et al.*, 2013; *Hayden et al.*, 2019; *Weitz et al.*, 2008; *Wiseman et al.*, 2008].

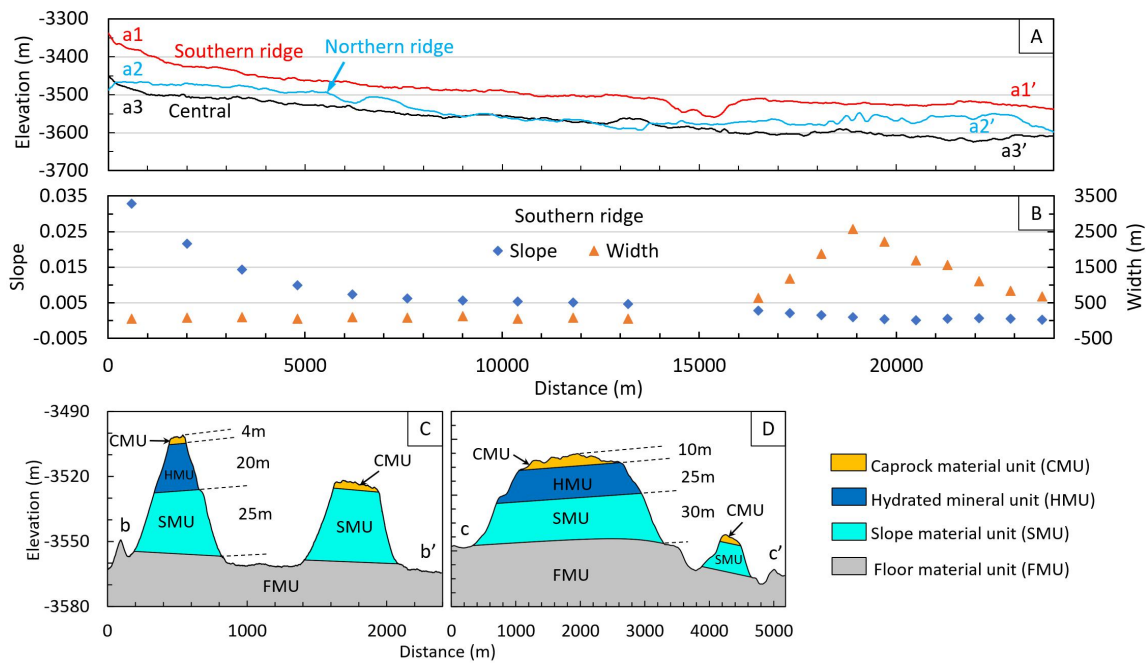

**Figure 4.** (a) Topographic profiles of the southern ridge, the northern ridge, and the trough in the middle. (b) Cross section profile of the ridges in Zone I. (c) Cross-sectional profile of the ridges in Zone II. (d) The stratigraphic profile shows the geologic units across the ridges: caprock material unit (CMU, 4 m thick), hydrated mineral unit (HMU, 20 m thick), slope material unit (SMU, 25 m thick) and the floor material unit (FMU).

**Discussion**

Several interpretations have been proposed for sinuous ridges on Mars, including lava tubes, eskers, inverted channels, and inverted channel belt [*Bernhardt et al.*, 2013; *Bernhardt et al.*, 2019; *Bleacher et al.*, 2017; *Davis et al.*, 2019; *Gallagher* and *Balme*, 2015; *Guidat et al.*, 2015; *Hayden et al.*, 2019; *Lefort et al.*, 2012; *Newsom et al.*, 2010; *Pain et al.*, 2007; *Ramsdale et al.*, 2015; *Williams et al.*, 2013; *Zealey*, 2009; *Zhao et al.*, 2017]. Lava tubes are mostly associated with volcanic activities and lava flows. Although the Alba shield volcano nearby could provide the lava source for the formation of lava tube, this model cannot well explain the origin of smectite-bearing clay minerals underneath the capping unit. Eskers are composed of sands and gravels transported by subglacial channel flow. Subglacial streams, consisting of glacial meltwater, will bring and deposit the sediments and form an esker. Smectite-bearing clay minerals could be formed during this process. However, eskers normally have sharp or rounded crests [*Butcher et al.*, 2020; *Kargel* and *Storm*, 1992; *Pain et al.*, 2007], contrary to our observations and there is little evidence for glacial activities within the topographic depression. Also, the caprock trends consistently downhill while an esker origin commonly produces greater topographic undulations [*Butcher et al.*, 2020].

The inverted channel hypothesis has been widely used to explain the sinuous ridges observed on Mars [e.g., *Williams et al.*, 2013]. In this model, deposits fill the geometry of the paleo-channel to form a series of sedimentary materials, and during diagenesis (e.g., cementation) a layer of caprock is formed which defines the erosion resistant materials on the riverbed [*Williams et al.*, 2013]. After further deposition and burial, the environment becomes arid and the former riverbed and its surrounding area would undergo weathering and erosion. Because the riverbed is covered with a layer of erosion-resisted caprock, it is much less eroded as compared to the surroundings. The riverbed will be raised and become a ridge [*Pain et al.*, 2007; *Zaki et al.*, 2018]. This model also faces some challenges, as the channel fills are commonly finer-grained deposits which may be less likely to be preserved to form inverted ridges [*Bhattacharya et al.*, 2016; *Bridge*, 2003; *Hayden et al.*, 2019; *Musial et al.*, 2012].

An alternative explanation of the inverted sinuous ridge observed in the study area is inverted channel belt (**Figure S1**). An inverted channel belt represents the exhumed landform reflecting the lateral migration of rivers and vertical aggradation of river deposits rather than the original channel geometry [*DiBiase et al.*, 2013; *Hayden et al.*, 2019; *Mohrig et al.*, 2000]. The repeated flooding and deposition will form a channel belt sandstone body that is wider than the width of the river itself. In addition, during the migration of the river, floodwater will overflow the riverbank and sediments will be transported to the floodplain. During exhumation, the elevation of the channel belt sandstone will be raised to form the ridge terrain due to the protection of the caprock and the channel fills may not be preserved, and late erosion would reduce the ridge width by lateral backwasting of the channel belt sandstone [*Hayden et al.*, 2019].

The sinuous ridges observed in our study area are consistent with the inverted channel belt model of *Hayden et al.* [2019]. A river floodplain and channel belt forms where the river meandering over a relatively flat area floods the area and deposits the river sediments on it. The topographic profiles of the ridges show that the river was flowing from west to east following the dichotomy slope (**Figure 4a**). The ridges (i.e., the southern ridge) at the more flattened downstream area (i.e., Zone II; flat-crest top is ~1000-3000 m wide) are much wider but less steeper than that at the upstream area (i.e., Zone I; flat-crest top is ~50-200 m wide) (**Figures 4b-d**), suggesting that they may represent channel-belt deposits rather than the channel itself. During the formation of channel belts in fluvial systems, sediments can be deposited by either lateral expansion or downstream translation [*Bridge*, 2003; *Ghinassi et al.*, 2016; *Willis* and *Tang*, 2010]. In areas with gentle slopes such as the downstream of our study area (**Figure 4b**), the channel belts could be wider due to more extensive lateral migration and aggradation, which is consistent with our observations (**Figure 2a**). The parallel northern ridge could also be an inverted channel belt. The two ridges do intersect, but at different topographic levels. Therefore, the two channel-belt deposits probably record rivers that formed at different times, with the lower-elevation channel-belt deposits (i.e., the northern ridge) forming first (see **Figure S2** for the preferred stratigraphic interpretation). By analogy to terrestrial river-and-floodplain deposits [*Bridge*, 2003], which record frequent channel avulsions, these two distinct deposits might record intervals in the history of a single river.

Layered smectite-bearing clay minerals are distributed along both sides of the southern ridge flanks at the same stratigraphic level (**Figure 2b**), which could correspond to the mudstone and sandstone of the floodplain. Indeed, the smectite-bearing clay minerals are not only distributed on the ridge slope, but also exposed in various settings as a continuous layer of phyllosilicates close to the ridge (**Figure 2A**), indicating the possible wide distribution of the clay minerals on the floodplain. These clay minerals could have been transported to their current location from the topographically higher surrounding plains, which would require the clay mineral-bearing source units for the detrital sediments [*Goudge et al.*, 2015; *Milliken* and *Bish*, 2010]. However, we surveyed the potential source regions to the west of our study area and did not find widespread exposures of clay minerals. Although it is possible that some local clay deposits exist in the source region but are concealed by dust, the absence of large outcrops suggest that the clay minerals are likely not detrital in origin, but instead authigenic, as is inferred for clays detected at Gale crater by the Curiosity rover. Alternatively, fine-grained materials such as ash or wind-blown sand/dust were transported to the area by wind during a period with less runoff and were then altered to form clay minerals by *in situ* pedogenesis [*Carter et al.*, 2015; *Le Deit et al.*, 2012; *Loizeau et al.*, 2018]. In this case, the rivers flowed only during brief periods and reworked the wind-blown materials. In Miyamoto crater, on the dichotomy near Arabia Terra, clay minerals have also been found to be associated with the sinuous ridges, and similarly to our site, these clay minerals are distributed along the slope of the ridges as well as on the crater floor [*Newsom et al.*, 2010; *Wiseman et al.*, 2008]. The clay minerals in Miyamoto crater were interpreted to be formed by either deposition or *in situ* alteration by fluvial activities. Sinuous ridges are widely distributed at the dichotomy boundary on Mars [*Guidat et al.*, 2015; *Lefort et al.*, 2012], and the processes to form the sinuous ridge - clay mineral association may not be uncommon at the dichotomy boundary. Future robotic missions, such as the ExoMars Rosalind Franklin rover, planned to land in the Oxia Planum region near the dichotomy [*Quantin-Nataf et al.*, 2020], may determine whether the origin of these clay minerals is detrital or authigenic.

The inverted channel belts and the associated clay minerals deposited on the channel belts and the floodplain identified in this study have important implications for the aqueous history of the region. The inverted ridges are located on mid-late Noachian terrain, around Mars' dichotomy boundary where valley networks terminate in deltas that may have formed along the coastline of an putative ancient ocean in northern lowlands [*Fawdon et al.*, 2018; *Parker et al.*, 1989; *Rodriguez et al.*, 2016]. A major open question is the duration of fluvial deposition at this proposed paleo-ocean shoreline. The thick clay-bearing layer (~ 20 m) on the flank of the 30 km long inverted channel belt and the widespread of clay minerals on the interpreted floodplain, stratigraphically bracketed by channel belt deposits, implies prolonged fluvial activity on early Mars. At long-term terrestrial floodplain aggradation rates, a thickness of ~30 m corresponds to a time interval of $1.5 \times 10^3 - 9 \times 10^4$ yrs [*Bridge* and *Leeder*, 1979]. This timescale could be much longer on Mars due to intermittency.

Our site is antipodal to the Gale-Aeolis-Zephyria region, therefore, evidence for prolonged fluvial activity [*Kite et al.*, 2015; *Kite et al.*, 2013b] is globally distributed. It has been debated whether the Noachian period on Mars is "warm and wet" or "cold and icy". The valley networks and open-basin lakes distributed on the southern highlands indicate abundant liquid water, suggesting a warm and wet scenario of the Mars climate in Noachian. However, climate models of Mars indicate that Mars could have been cold and icy in the Noachian and the fluvial landforms on Mars' surface may have been formed by flows of liquid water supplied by transient melting of snow or ice deposits [*Wordsworth et al.*, 2013; *Wordsworth et al.*, 2015]. If this is the case, the inverted channel belts and the floodplain clays and the implied persistent fluvial activity in our study area indicate that episodic warming events probably triggered by volcanism, impacts, and/or variations in planetary obliquity have lasted over at least tens of thousands of years in the Noachian period of Mars.

## Conclusion

We report the identification of a series of inverted ridges to the east of Tempe Terra on Mars. These inverted ridges are interpreted to have formed by exhumation of the channel belts and overbank deposits that were formed from meander rivers over significant geologic time, which may shed light on the complex history of erosion and river activity close to the shoreline of the proposed Northern paleo-ocean on Mars. The widespread smectite-bearing clay minerals on the flank of the inverted channel belt and the interpreted floodplain imply prolong fluvial activity on Mars, potentially for tens of thousands of years (or longer if intermittency is taken into account), and our study area has experienced episodic warming events that have lasted over deep time in Noachian period of Mars.

## Acknowledgements

We thank Caleb Fassett at NASA Marshall Space Flight Center for providing the CTX DEM and Marssi for the help with generating the HiRISE DEM products. We thank the CRISM, HiRISE, CTX, MOLA teams for making Martian data available in this project. All the data used in this research are available through the Planetary Data System (https://pds-geosciences.wustl.edu/).